\documentclass[aps,twocolumn,showpacs,showkeys,amsmath,amssymb,nofootinbib,superscriptaddress,prc,floatfix]{revtex4-1}
\usepackage{graphicx}
\usepackage{subfigure}
\usepackage{bm}

\makeatletter
\newcommand\erfc{\mathop{\operator@font erfc}\nolimits}
\def\slashchar#1{\setbox0=\hbox{$#1$}
   \dimen0=\wd0 \setbox1=\hbox{/} \dimen1=\wd1
   \ifdim\dimen0>\dimen1 \rlap{\hbox to \dimen0{\hfil/\hfil}} #1
   \else  \rlap{\hbox to \dimen1{\hfil$#1$\hfil}} / \fi}

\begin{document}
\title{Indications of early thermalization in relativistic heavy-ion collisions}
\author{Piotr Bo\.zek}
\affiliation{
Institute of Physics, Rzesz\'ow University, 
PL-35959 Rzesz\'ow, Poland}
\affiliation{The H. Niewodnicza\'nski Institute of Nuclear Physics,
PL-31342 Krak\'ow, Poland} 
\author{Iwona Wyskiel-Piekarska}
\affiliation{The H. Niewodnicza\'nski Institute of Nuclear Physics,
PL-31342 Krak\'ow, Poland}
\date{\today}

\begin{abstract}
The directed flow of particles emitted from the  fireball
created in a heavy-ion collision is shown to be
 a very sensitive measure of the
 pressure equilibration in the first $1$ fm/c   of the evolution. Performing a 
3+1 dimensional (3+1D) relativistic hydrodynamic calculation with nonequilibrated
 longitudinal and transverse pressures, we show that the directed flow is
 strongly reduced if the pressure imbalance survives for even a short time.
Transverse momentum spectra, elliptic 
flow and interferometry correlation radii are not very sensitive to this early 
pressure anisotropy. Comparison with  the data  points toward a short
 equilibration time of the order of $0.25$fm/c or less.
\end{abstract}

\pacs{25.75.Ld, 24.10Nz, 24.10Pa}

\keywords{relativistic 
heavy-ion collisions,  hydrodynamic model, collective flow, directed flow}

\maketitle

\section{Introduction}

Accumulated experimental observations
 from  heavy-ion collisions at 
  the BNL 
Relativistic Heavy Ion Collider  indicate that a fireball of
dense and hot matter is formed in the course of the collision
\cite{Arsene:2004fa,*Back:2004je,*Adams:2005dq,*Adcox:2004mh}.  
The medium behaves as an almost perfect, thermalized 
 fluid, with very small viscosity.
Relativistic hydrodynamic models  describe the 
 experimental data for transverse momentum  ($p_\perp$)  spectra,
Hanbury Brown-Twiss  (HBT) correlations radii and elliptic flow of particles
\cite{Kolb:2003dz,*Huovinen:2006jp,*Hirano:2008aj,*Ollitrault:2010tn,Broniowski:2008vp,*Pratt:2008qv,Bozek:2009ty}. 
Small
 deviations from local equilibrium due to
 large velocity gradients in the expansion have been quantified within 
the relativistic viscous hydrodynamics 
\cite{IS,Teaney:2003kp,Song:2007ux,Dusling:2009df,Chaudhuri:2006jd,Dusling:2007gi,Romatschke:2009im,Teaney:2009qa,Luzum:2008cw,Bozek:2009dw,Schenke:2010rr}. 

A separate, still unresolved  question is how fast the initial, 
almost thermally equilibrated fluid is formed. 
 If this time 
scale is very short ($<0.5$fm/c), it would indicate that a strongly coupled
 system is formed in the collision. On the other hand, a longer equilibration
 time is compatible with weakly coupled perturbative QCD. 
Microscopic mechanisms 
responsible for the creation of the dense fireball are the subject of intense
 theoretical studies. Scenarios such as the development of instabilities of color fields, 
evolution of parton distributions in color fields, 
collisional equilibration,
 strong coupling solutions  are proposed \cite{Mrowczynski:2005ki,Rebhan:2008uj,Bjoraker:2000cf,Xu:2004mz,Chesler:2009cy,Beuf:2009cx}.
Models predict a larger value for the transverse $P_\perp$ than the 
longitudinal 
$P_\parallel$ pressure in the early stage. After some isotropization time 
$\tau_{izo}$, the 
two pressures become similar. The isotropization of the pressure 
is a necessary, but not a sufficient condition for equilibrium.  However, in
 the following, we 
understand the equilibration time as the time when the pressure becomes 
(almost)
isotropic, from that moment ideal (or more correctly, viscous) 
hydrodynamics applies. This means that we follow the natural assumption
 that the pressure
 equilibration is related to some microscopic processes driving the system 
toward equilibrium; in that case the time for the pressure isotropization
 can be used as an estimate for the equilibration time.

\section{Nonisotropic pressure}

The onset of the collective expansion in heavy-ion collisions 
can be separated into
several stages. First, interacting, dense matter must be created after the 
initial nucleon collisions. Such an interacting system, can be 
effectively described using an energy-momentum tensor. For a system 
in equilibrium, the energy-momentum tensor 
 has the form
\begin{equation}
T^{\mu\nu}=(\epsilon+P_{eq})u^\mu u^\nu - P_{eq} g^{\mu \nu} \ ,
\end{equation}
where $u^\mu$ the fluid velocity, 
$\epsilon$ the energy density, and $P_{eq}$ the pressure. The
creation of the interacting matter, close to equilibrium, does not mean 
that the elementary degrees of freedom are quasiparticles. The strongly 
interacting plasma is an almost perfect fluid 
\cite{Luzum:2008cw,Song:2008hj,Bozek:2009dw}. The small value of the 
shear viscosity
indicates that the system is very far from the limit of a kinetic description
 \cite{Kovtun:2004de,Danielewicz:1984ww,Liao:2009gb}. 
When the nonequilibrium effects represent only a correction to the ideal
 fluid picture, the viscous hydrodynamics can be applied \cite{IS}.  Second
 order viscous hydrodynamics 
 applies if the product of the viscosity coefficient 
times the velocity gradients is relatively
small and if the initial pressures are close to equilibrium. In this paper 
we study the dynamics in the 
 early stage, when the second assumption is not valid.

 There is no realistic model of the
equilibration that could be applied to heavy-ion phenomenology. In a 
short-living and strongly interacting system, quasiparticles cannot be 
formed. This causes problems for the application of kinetic models 
to the early phase \cite{Xu:2004mz,Martinez:2010sc,*Martinez:2010sd}. Field theory approaches
use simplified models or geometries
\cite{Rebhan:2008uj,Bjoraker:2000cf,Chesler:2009cy,Beuf:2009cx,Dusling:2010rm}.
Formally, relativistic viscous hydrodynamics breaks down at short time scales,
 as stress corrections to the energy-momentum tensor are dominant.
In this paper, we do not develop a microscopic model
of the early evolution. Instead, we
 focus on possible observable consequences of the
early nonequilibrium evolution.
We assume that the nonequilibrium effects manifest themselves 
in the energy-momentum tensor as anisotropic pressure
\begin{equation}
T^{\mu\nu}=\left( \begin{array}{cccc} \epsilon & 0 & 0 &0\\
0& P_{eq}+\pi/2 & 0 & 0\\
0& 0 & P_{eq}+\pi/2 & 0 \\
0 & 0 & 0 & P_{eq}-\pi 
 \end{array}\right) \ . 
\label{eq:tmunu}
\end{equation}
In the early evolution, one expects that the
 transverse pressure  is larger and the longitudinal 
pressure  is smaller than the equilibrium one $P_{eq}$.
This form of the energy-momentum tensor
 appears in the viscous hydrodynamics with Bjorken flow \cite{Teaney:2003kp}.
Similar correction to the longitudinal and transverse pressures are expected
in a solution of strongly interacting systems or in the presence of 
classical color fields  \cite{Chesler:2009cy,Beuf:2009cx,Vredevoogd:2008id}.
If a theory could be applied to the far from equilibrium evolution, the 
 correction to the pressure $\pi$ would be a solution 
of dynamical equations. Our goal is to study the effects of the possible 
presence of the nonequilibrium correction on the fluid dynamics
 and not to calculate the correction from an underlying theory. For that purpose
 it is sufficient to assume a time dependence of the stress correction $\pi$
 that describes the approach to equilibrium and use it to calculate 
its effect on observables in heavy ion collisions. The early equilibration
 means that stress corrections decrease with time, reaching  zero at 
 equilibrium.
 In a more realistic scenario the stress correction is not vanishing 
at large times, but should approach the value given by the 
shear viscosity effects. We do not take this into account in this first
 estimate, as we are interested in the very early stage of the dynamics.

The time scale when the pressure  anisotropy decays
 is an estimate of the isotropization time, which for the bulk dynamics
 is equivalent to the thermalization time.
A question of primary importance is the possibility of getting an 
experimental estimate of the equilibration time. 
Promising observables
 sensitive to the early dynamics are the dilepton and 
photon emissions
\cite{Mauricio:2007vz,*Schenke:2006yp,*Dusling:2008xj,*Dusling:2009bc,*Bhattacharya:2008mv}. 
The difficulties of 
 such calculations are the uncertainties in the higher order corrections to 
the emission rates and a large 
background from other sources of dileptons and photons.
Other possibilities \cite{Bozek:2007di,Florkowski:2010cf,*Ryblewski:2010bs,Martinez:2010sc} 
that have been explored study the effects of early off-equilibrium 
effects in the energy-momentum tensor on the hydrodynamic evolution.

Observables, such as $p_\perp$ spectra,
elliptic flow, or HBT radii are sensitive 
 to the transverse flow profile at  the freeze-out.
The transverse flow  is built up from the acceleration in the 
transverse direction, during the whole hydrodynamic evolution.
In a boost invariant geometry the reduced longitudinal pressure acts only 
by changing the cooling rate. After fixing the final entropy 
per unit rapidity in all the scenarios, 
involving different longitudinal pressures, 
one obtains 
similar results \cite{Broniowski:2008qk,Vredevoogd:2008id}. 
In the following we show,  in a realistic $3+1$D hydrodynamic 
simulation, that the $p_\perp$ spectra, HBT radii, 
and elliptic flow are indeed 
not sensitive to the early pressure  anisotropy. 

One could argue that the
longitudinal expansion itself 
is sensitive to the reduced longitudinal pressure. 
However,  the experimentally observed rapidity distributions can be reproduced 
by many scenarios with reduced $P_\parallel$, simply 
   using different initial conditions \cite{Bozek:2007qt}.
With no a priori knowledge about the initial energy density, no
conclusions can be drawn about how much and for how long 
the longitudinal pressure is reduced.

The initial energy density $\epsilon(\eta_\parallel,x,y)$ 
 for the $3+1$D hydrodynamic
 evolution
in the space-time rapidity $\eta_\parallel$ 
and the transverse plane $(x,y)$ is taken  in the form
\begin{eqnarray}
\epsilon & \propto&\left[ \left( \frac{(\eta_m+\eta_\parallel)N_+
+(\eta_m-\eta_\parallel)N_-}{\eta_m(N_++N_-)}\right)
  \frac{1-\alpha}{2}\rho_{part}\right. \nonumber \\
& &\left. +\alpha \rho_{bin}\right] f(\eta_\parallel)
\label{eq:eini}
\end{eqnarray}
where the density in the transverse plane is proportional to a combination of 
participant nucleon $\rho_{part}=N_++N_-$ and binary collision 
$\rho_{bin}$ densities. 
On the right hand side of Eq. (\ref{eq:eini}), 
only the dependence on $\eta_\parallel$ is shown explicitly. 
The dependence on the position in the transverse plane comes
 from the densities of the right and left
 going participant 
nucleons $N_\pm(x,y)$, calculated from the Glauber model. 
The parameters and the longitudinal profile $f(\eta_\parallel)$ 
are adjusted to reproduce the  experimental spectra  
 \cite{Bozek:2009ty}.

The factor 
\begin{equation}
\frac{(\eta_m+\eta_\parallel)N_++(\eta_m-\eta_\parallel)N_-}{\eta_m(N_++N_-)}
\end{equation}
in the initial energy density (\ref{eq:eini}) is based on the assumption that 
forward going participant nucleons with density $N_+$
 emit particles predominantly 
in the forward hemisphere, with  the reverse for the backward going particles. 
Such a distribution is expected in bremsstrahlung emission, and has been 
extracted from the analysis of the data in asymmetric collisions 
\cite{Bialas:2004su,*Adil:2005qn}. The beam rapidity 
 $\eta_m=\ln(\sqrt{s}/m_p)$
 is taken as the rapidity of the
emitting charge. 
The effect 
of this modification of the initial density is to cause
 a tilt of the initial source away from the collision axis (see Fig. 3 in 
\cite{Bozek:2010bi}).
In (\ref{eq:eini}) only the density of participant nucleons is tilted. If we assume that the density of binary collisions is tilted as well, we have
\begin{eqnarray}
\epsilon & \propto&\left[ \left( \frac{(\eta_m+\eta_\parallel)N_+
+(\eta_m-\eta_\parallel)N_-}{\eta_m(N_++N_-)}\right) \right. \nonumber \\
& &\left. \left(\frac{1-\alpha}{2}\rho_{part} +\alpha \rho_{bin}\right) \right] f(\eta_\parallel) \ ,
\label{eq:einilarge}
\end{eqnarray}
which we call large tilt initial conditions. 
The variation in the magnitude of the initial  tilt between formulas (\ref{eq:eini}) and (\ref{eq:einilarge})
 is a measure of the uncertainty of the
 model.

Using the tilted initial conditions, 
 the coefficient $v_1$ of the directed flow 
\cite{Abelev:2008jga} of particles 
emitted in Au-Au collisions 
at $\sqrt{s}=200$GeV \cite{Bozek:2010bi} has been described
within the ideal fluid hydrodynamic model.
Two important characteristics 
of this mechanism are of relevance for the study
 of the  early isotropization. First, the directed flow is generated very 
early in the dynamics, mainly in the first $1$ fm/c.  Second, the formation 
of the directed flow requires a simultaneous acceleration of fluid elements
 by the transverse and longitudinal pressures \cite{Bozek:2010bi}.

The mechanism generating 
the  directed collective flow from a tilted initial
 source can be understood when considering the acceleration equations of 
 relativistic hydrodynamics. For small initial times 
(small transverse velocities $v_x, v_y$) the accelerations 
in the transverse $x$  and longitudinal $\eta_\parallel$ 
directions take the form
\begin{eqnarray}
{\partial_\tau v_x}=&&-\frac{\partial_x P_\perp}{\epsilon + P} \ , 
\label{eq:ac1} \\
{\partial_\tau Y}=&&-\frac{\partial_{\eta_\parallel} P_\parallel}{\tau(\epsilon + P)} \ ,
\label{eq:ac2} \end{eqnarray}
where $Y=\frac{1}{2}\ln\left((1+v_z)(1-v_z)\right)$
is the rapidity of the fluid element. At freeze-out, a fluid 
element of rapidity $Y$ and velocity $v_x$ emits particles with
 a thermal smearing around these velocities. A negative directed flow 
$v_1<0$ for positive pseudorapidities $\eta>0$, means that most of the
emission happens in fluid elements with negative correlation between 
$v_x$ and $Y$. In a tilted source the early acceleration of the fluid 
occurs predominantly with  anticorrelated signs in the 
$v_x$ (\ref{eq:ac1}) and $Y$ (\ref{eq:ac2}) directions.
If in the early stage the longitudinal pressure $P_\parallel$ 
is significantly reduced, the directed flow is not  generated.
The sensitivity of the directed flow to the simultaneous action 
of the transverse and longitudinal pressures, makes it a preferred 
observable to measure the degree of pressure anisotropy.

\section{Results}

\begin{figure}
\includegraphics[width=.48\textwidth]{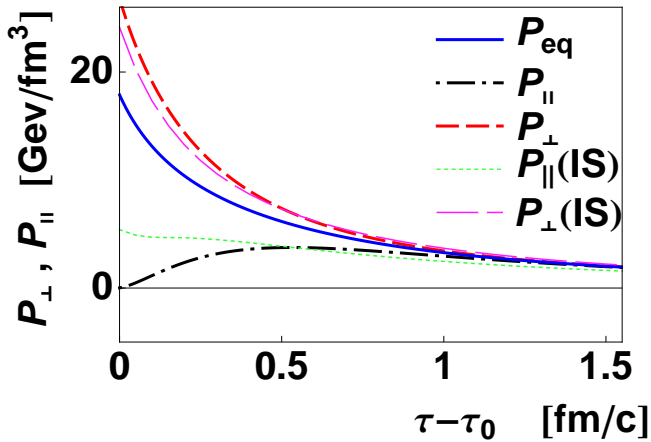}
\caption{(Color online) Time evolution of the longitudinal, 
transverse and equilibrium  pressures (dash-dotted, dashed and 
solid lines respectively) at the center of the fireball in a
 central Au-Au collision, with 
 $P_L(\tau_0)=0$ and $\tau_{iso}=0.25$fm/c.
 The long-dashed and dotted lines represent the transverse and longitudinal pressures   in  viscous hydrodynamics  with $\eta/s=1/4\pi$. }
\label{fig:peq}
\end{figure}

\begin{figure}
\includegraphics[width=.48\textwidth]{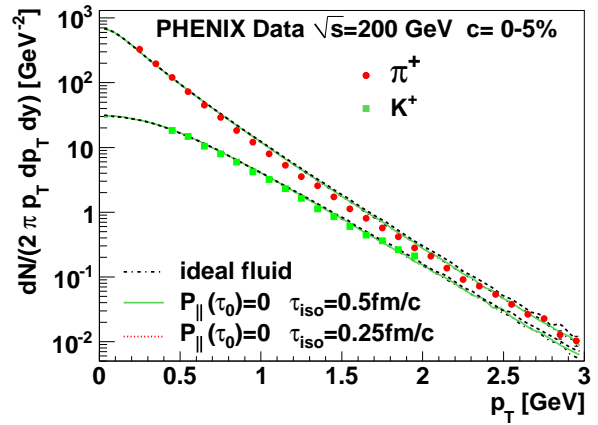}
\caption{(Color online) Transverse momentum spectra for $\pi^+$ and $K^{+}$. The dash-dotted, solid and dotted  lines represent
 the hydrodynamic results with $\tau_{iso}=0$, $0.25$, and $0.5$fm/c respectively; data are from the 
PHENIX Collaboration \cite{Adler:2003cb}.}
\label{fig:sp}
\end{figure}

We perform numerical simulations of the  $3+1$D hydrodynamics with 
anisotropic pressures. A phenomenological 
correction is added (\ref{eq:tmunu})  to the ideal fluid energy-momentum tensor
with
\begin{equation}
\pi(\tau,x,y,\eta_\parallel)=P_{eq}(\tau_0,x,y,\eta_\parallel)
 e^{(\tau_0-\tau)/\tau_{iso}}\ .
\label{eq:anip}
\end{equation}
The pressure anisotropy makes the longitudinal 
pressure $P_\parallel=P_{eq}-\pi$ zero initially, and the 
anisotropy decays exponentially with a relaxation  time $\tau_{iso}$.
A similar ansatz has been used in boost invariant calculation of 
the effect of the initial dissipation on the transverse expansion 
\cite{Bozek:2007di}. Other parameterizations of the form of the 
initial pressure anisotropy have been used 
\cite{Mauricio:2007vz,*Schenke:2006yp,*Dusling:2008xj,Ryblewski:2010bs}.
 The goal of this paper is to propose
a generic signature of the presence of the early anisotropy of the pressure,
 that can be used for any time dependence of the stress tensor correction, e.g.
from microscopic models, as long as the far from equilibrium dynamics 
in the microscopic theory would imply, in the hydrodynamic limit,  a pressure
anisotropy.

 The actual form and time dependence of 
the pressure anisotropy are not known from first principles. One 
expects that the large initial anisotropy decays and becomes small
 in the latter
 stages of the collision, as the shear viscosity coefficient of
 the matter formed in nuclear collisions is estimated to be
 small \cite{Romatschke:2009im,Teaney:2009qa}. 
 Similar relaxation time terms appear in the 
second order relativistic viscous hydrodynamics \cite{IS}, but the present 
investigation does not rely 
 on the assumption of viscous hydrodynamics, which is
 not applicable at this very early stage, where the dissipative 
correction to the pressure is of the same order as the pressure itself.

The assumed form (\ref{eq:anip}) is a phenomenological ansatz used to test the
sensitivity of  different observables on the early pressure anisotropy.
The time scale in this parameterization is given by the isotropization time
 $\tau_{iso}$.
Another parameter is the initial value of the pressure anisotropy 
at the beginning of the collective expansion. In Eq. \ref{eq:anip} we take 
$\pi(\tau_0)=P_{eq}(\tau_0)$, which gives zero longitudinal
 pressure at the beginning. The initial pressure is even less
constrained than the time dependence of $\pi$, as it is defined 
by the energy deposition processes in the collision before the formation
of the dense fireball. 
 We have checked that qualitatively similar results are obtained,  if the initial longitudinal pressure is varied between 
$P_L(\tau_0)=P_{eq}/2$  and $P_L(\tau_0)=-P_{eq}/2$.
Initial conditions with zero or negative initial longitudinal
 pressure cannot be described 
in a kinetic theory or as viscous corrections to the ideal hydrodynamics.

The hydrodynamic equation are obtained boosting the 
 energy-momentum tensor (\ref{eq:tmunu})  by the fluid velocity $u^\mu$
\begin{equation}
T^{\mu\nu}=(\epsilon+P_{eq}+\frac{\pi}{2})u^\mu u^\nu-(P_{eq}+\frac{\pi}{2})g^{\mu\nu}-\frac{3\pi}{2} v^\mu v^\nu \ \ , 
\end{equation}
with 
\begin{equation}
u^\mu=(\gamma \cosh Y,u_x ,u_y, \gamma \sinh Y )\ , 
\end{equation}
and
\begin{widetext}
\begin{equation}
v^\mu=\left( \gamma \sinh Y , \frac{\sinh Y u_x(\gamma^2\cosh Y -\gamma)}{\sinh Y +\cosh Y (u_x^2+u_y^2)}, \frac{\sinh Y u_y(\gamma^2\cosh Y -\gamma)}{\sinh Y +\cosh Y (u_x^2+u_y^2)},  \frac{u_x^2+u_y^2+\sinh^2 Y \cosh Y \gamma^3}{\sinh Y +\cosh Y (u_x^2+u_y^2)}\right) \ , 
\end{equation}
\end{widetext}
where  $\gamma=\sqrt{1+u_x^2+u_y^2}$, $u_\mu v^\mu=0$. 
The resulting long hydrodynamic equations
\begin{equation}
\partial_\mu T^{\mu\nu}=0
\end{equation}
 are not  presented explicitly. We solve them numerically 
with initial conditions with the two different initial tilts of the source
 (\ref{eq:eini}) and (\ref{eq:einilarge}).
 The  initial time 
$\tau_0=0.25$fm/c is chosen. The small value for the time at which 
the transverse expansion begins is favored in order to 
reproduce the experimental HBT data \cite{Broniowski:2008vp}.
At the  freeze-out temperature of  $150$MeV, particles are emitted 
and resonance decays are performed using the event generator THERMINATOR
\cite{Kisiel:2005hn}. In a kinetic picture of the nonequilibrium pressure,
 the momentum distributions of particles have nonequilibrium corrections
 \cite{Teaney:2003kp}. Those modifications of the momentum distribution 
should be taken into account in the Cooper-Frye formula at the freeze-out. 
Such a kinetic picture of the dense matter at the very early stage is 
questionable, and is not used in the present investigation. The integration
 of the particle emission over the freeze-out hypersurface is taken for 
$\tau\ge 1 $fm/c. In our calculation, 
the nonequilibrium correction are reduced after that time 
and the equilibrium distributions are used in the Cooper-Frye formula.
We have checked that including the emission, using equilibrium distributions,
 for $\tau<1$~fm/c  
modifies the results by less than $3$\%.
In simulations with an initial
 pressure anisotropy,  entropy is generated. 
The initial density must be rescaled 
to take this effect into account. The relative equilibrium
 entropy production is $29\%$ and $61\%$ 
for $\tau_{iso}=0.25$ and $0.5$fm/c respectively, very close to the 
value estimated from a formula  valid in the Bjorken scaling expansion
 \cite{Bozek:2007di}.

\begin{figure}
\includegraphics[width=.4\textwidth]{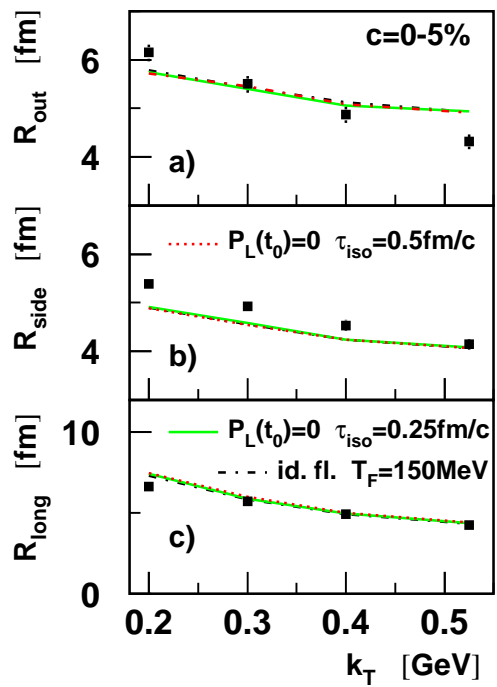}
\caption{(Color online) 
 STAR Collaboration data \cite{Adams:2004yc} (squares)  for the HBT radii  
($R_{out}$, $R_{side}$ and $R_{long}$ in panels a), b) and c) respectively)
 compared to the results of  hydrodynamic simulations  
(same lines as in Fig. \ref{fig:sp}). }
\label{fig:hbt}
\end{figure}

In Fig. \ref{fig:peq} is shown the time dependence of 
 the longitudinal and transverse pressures. 
The pressures equilibrate fast to the isotropic value.
 Figs. \ref{fig:sp} and \ref{fig:hbt} show results for the 
transverse momentum spectra and HBT radii 
for Au-Au collisions of centrality $0-5\%$. For the three cases considered~:
 isotropization with $\tau_{iso}=0$ (ideal fluid), $0.25$ and $0.5$fm/c,
the $p_\perp$ spectra and the HBT  radii are indistinguishable. 
 This confirms the existence of a universality in the 
transverse flow for different longitudinal pressures \cite{Vredevoogd:2008id}.
The reduced longitudinal pressure  in the isotropization scenario, 
implies a slower cooling, but renormalizing the initial conditions to obtain 
the same final multiplicity cancels this effect. The source lives
 for the same total time and the same transverse flow is generated in
 all the cases considered.

\begin{figure}
\includegraphics[width=.48\textwidth]{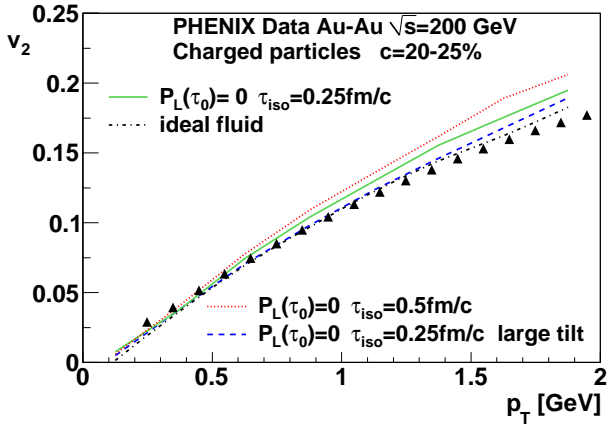}
\caption{(Color online) The elliptic flow coefficient $v_2$ of charged particles
from the PHENIX Collaboration  \cite{Adare:2010ux} (triangles) for  
 centrality $20-25\%$ compared to hydrodynamic calculations; 
the  dashed line is for $\tau_{iso}=0.25$fm/c and the maximal tilt of the 
source (Eq. \ref{eq:einilarge}) and other lines are as in Fig. \ref{fig:sp}.
}
\label{fig:v2}
\end{figure}

For  semi-central collisions
 azimuthally asymmetric emission can be tested. 
 The elliptic flow coefficient $v_2$ (Fig. \ref{fig:v2})
 is very similar for the different isotropization times tested.
 Again, we find that the elliptic flow is not
 an observable sensitive to the early pressure anisotropy.

\begin{figure}
\includegraphics[width=.48\textwidth]{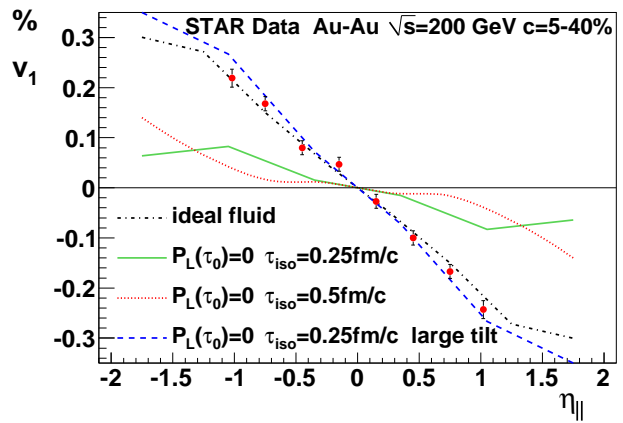}
\caption{(Color online) The directed flow from hydrodynamic calculations 
(same lines as in Fig. 
\ref{fig:v2}) compared to STAR Collaboration data 
\cite{Abelev:2008jga}.}
\label{fig:v1por}
\end{figure}

On the other hand, the directed flow  varies very much 
depending on the longitudinal pressure active 
in the early expansion.   Fig. \ref{fig:v1por} presents $v_1$
 for charged particles as a function of pseudorapidity. When 
the longitudinal pressure is reduced, the directed flow is not generated 
with enough strength. It demonstrates  that the longitudinal 
acceleration must be active (Eq. \ref{eq:ac2}) very early to 
generate enough directed flow.

To estimate the thermalization time $\tau_{iso}$,  the
initial deformation of the source must be known. We use two extreme assumptions
for the value of the tilt. For the smaller tilt (Eq. \ref{eq:eini}), 
the experimental data is described using $\tau_{iso}=0$ (ideal fluid).
The expansion of the source with the larger tilt (Eq. \ref{eq:einilarge})
 is compatible with the
data if the longitudinal pressure is retarded by $0.25$fm/c with respect to the 
transverse pressure. The conclusion of this analysis is that the isotropization time
of the pressure is smaller than $0.25$fm/c. This time 
can be treated as an effective thermalization time of the medium.

%

The question arises whether such a small degree of pressure anisotropy 
is compatible with viscous corrections to the ideal fluid flow, which 
is the minimal mechanism generating nonequilibrium corrections to the 
energy-momentum tensor. Full $3+1$D viscous 
hydrodynamic calculations of the directed flow are not yet available. For the 
Bjorken flow at the early stage, shear viscosity  corrections have the
same  form as in Eq. (\ref{eq:tmunu}) and
the correction to the pressure $\pi$ is 
the solution of a dynamical equation \cite{IS}. In Fig. \ref{fig:peq} 
are shown the longitudinal (dotted line) and transverse (long-dashed line)
 pressures resulting from the 
action of 
the shear viscosity, corresponding to the strong coupling limit
($\eta/s=1/4\pi$)
\cite{Kovtun:2004de}, with the 
 Navier-Stokes initial value of the
stress correction $\pi=4\eta/3\tau_0$. The pressure anisotropy 
 from such a small shear viscosity  is compatible with  our limits 
on the pressure anisotropy. It indicates that the shear 
viscosity  in the dense matter at 
the early stage of the collision is close to the strong coupling limit.
A different issue is the role of shear and bulk viscosities in the latter 
expansion. In particular, the viscosity and dissipative effects in the hadronic 
rescattering are known to influence significantly the final elliptic flow
\cite{Hirano:2005xf,*Hirano:2007xd}.

\section{Conclusions}

We propose to measure the thermalization time in the early stage of 
a heavy-ion collision using the 
directed flow of particles. 
We demonstrate in explicit hydrodynamic calculations that the directed flow is 
significantly reduced in the presence of even 
a very short pressure anisotropy. The directed flow observable is unique,
 as it is sensitive simultaneously to the transverse and longitudinal 
pressures. Moreover, the directed flow is generated early in the expansion.
Hydrodynamic calculations indicate that
observables such as the  $p_\perp$ spectra, HBT radii and elliptic
 flow are sensitive to the whole evolution of the fireball, and feel 
the action of the transverse flow only. A short reduction of the longitudinal
 pressure does not influence these transverse flow observables.

Using  the initial  fireball densities calculated in the Glauber model,
 we estimate that the thermalization time 
is smaller than $0.25$fm/c. Such a small value of the 
delay for the appearance of the 
longitudinal pressure indicates that the system is strongly coupled. 
For small deviations from equilibrium, the AdS/CFT result for the 
relaxation time  is 
$\tau_{\pi}=({1- \ln 2})/({6 \pi T})\simeq 0.02$~fm/c
\cite{Heller:2007qt}, which is $\simeq 30$ times smaller than the value 
for the massless Boltzmann gas. 
Our result also points toward a  small  shear viscosity in the 
dense plasma, as otherwise the longitudinal pressure would be significantly 
reduced. We note that the 
 directed flow could 
serve as a sensitive constraint for microscopic models of the 
initial equilibration \cite{Rebhan:2008uj,Bjoraker:2000cf,Xu:2004mz,Chesler:2009cy,Beuf:2009cx,Dusling:2010rm,Schenke:2010rr,Florkowski:2010cf,Martinez:2010sc};
 with
 approaches based on  field theory solutions \cite{Chesler:2009cy,Beuf:2009cx}
 or kinetic theory \cite{Xu:2004mz,Martinez:2010sc}
 being more general than our parameterization, 
as they describe both the far from equilibrium dynamics and the
 near equilibrium viscous hydrodynamics.

The work is supported  by the
Polish Ministry of Science and Higher Education 
grant No.  N N202 263438.

\bibliography{../hydr}

\begin{thebibliography}{10}%
\makeatletter
\providecommand \@ifxundefined [1]{%
 \ifx #1\undefined \expandafter \@firstoftwo
 \else \expandafter \@secondoftwo
\fi
}%
\providecommand \@ifnum [1]{%
 \ifnum #1\expandafter \@firstoftwo
 \else \expandafter \@secondoftwo
\fi
}%
\providecommand \enquote [1]{``#1''}%
\providecommand \bibnamefont  [1]{#1}%
\providecommand \bibfnamefont [1]{#1}%
\providecommand \citenamefont [1]{#1}%
\providecommand\href[0]{\@sanitize\@href}%
\providecommand\@href[1]{\endgroup\@@startlink{#1}\endgroup\@@href}%
\providecommand\@@href[1]{#1\@@endlink}%
\providecommand \@sanitize [0]{\begingroup\catcode`\&12\catcode`\#12\relax}%
\@ifxundefined \pdfoutput {\@firstoftwo}{%
 \@ifnum{\z@=\pdfoutput}{\@firstoftwo}{\@secondoftwo}%
}{%
 \providecommand\@@startlink[1]{\leavevmode\special{html:<a href="#1">}}%
 \providecommand\@@endlink[0]{\special{html:</a>}}%
}{%
 \providecommand\@@startlink[1]{%
  \leavevmode
  \pdfstartlink
   attr{/Border[0 0 1 ]/H/I/C[0 1 1]}%
   user{/Subtype/Link/A<</Type/Action/S/URI/URI(#1)>>}%
  \relax
 }%
 \providecommand\@@endlink[0]{\pdfendlink}%
}%
\providecommand \url  [0]{\begingroup\@sanitize \@url }%
\providecommand \@url [1]{\endgroup\@href {#1}{\urlprefix}}%
\providecommand \urlprefix [0]{URL }%
\providecommand \Eprint[0]{\href }%
\@ifxundefined \urlstyle {%
  \providecommand \doi [1]{doi:\discretionary{}{}{}#1}%
}{%
  \providecommand \doi [0]{doi:\discretionary{}{}{}\begingroup
  \urlstyle{rm}\Url }%
}%
\providecommand \doibase [0]{http://dx.doi.org/}%
\providecommand \Doi[1]{\href{\doibase#1}}%
\providecommand \bibAnnote [3]{%
  \BibitemShut{#1}%
  \begin{quotation}\noindent
    \textsc{Key:}\ #2\\\textsc{Annotation:}\ #3%
  \end{quotation}%
}%
\providecommand \bibAnnoteFile [2]{%
  \IfFileExists{#2}{\bibAnnote {#1} {#2} {\input{#2}}}{}%
}%
\providecommand \typeout [0]{\immediate \write \m@ne }%
\providecommand \selectlanguage [0]{\@gobble}%
\providecommand \bibinfo [0]{\@secondoftwo}%
\providecommand \bibfield [0]{\@secondoftwo}%
\providecommand \translation [1]{[#1]}%
\providecommand \BibitemOpen[0]{}%
\providecommand \bibitemStop [0]{}%
\providecommand \bibitemNoStop [0]{.\EOS\space}%
\providecommand \EOS [0]{\spacefactor3000\relax}%
\providecommand \BibitemShut [1]{\csname bibitem#1\endcsname}%
\bibitem{Arsene:2004fa}%
  \BibitemOpen
  \bibfield{author}{%
  \bibinfo {author} {\bibfnamefont{I.}~\bibnamefont{Arsene}} \emph{et~al.}
  (\bibinfo {collaboration} {BRAHMS}),\ }%
  \bibfield{journal}{%
  \Doi{10.1016/j.nuclphysa.2005.02.130}{\bibinfo {journal} {Nucl. Phys.}}\ }%
  \textbf{\bibinfo {volume} {A757}},\ \bibinfo {pages} {1} (\bibinfo {year}
  {2005})%
  \bibAnnoteFile{NoStop}{Arsene:2004fa}%
\bibitem{Back:2004je}%
  \BibitemOpen
  \bibfield{author}{%
  \bibinfo {author} {\bibfnamefont{B.~B.}\ \bibnamefont{Back}} \emph{et~al.}
  (\bibinfo {collaboration} {PHOBOS}),\ }%
  \bibfield{journal}{%
  \Doi{10.1016/j.nuclphysa.2005.03.084}{\bibinfo {journal} {Nucl. Phys.}}\ }%
  \textbf{\bibinfo {volume} {A757}},\ \bibinfo {pages} {28} (\bibinfo {year}
  {2005})%
  \bibAnnoteFile{NoStop}{Back:2004je}%
\bibitem{Adams:2005dq}%
  \BibitemOpen
  \bibfield{author}{%
  \bibinfo {author} {\bibfnamefont{J.}~\bibnamefont{Adams}} \emph{et~al.}
  (\bibinfo {collaboration} {STAR}),\ }%
  \bibfield{journal}{%
  \Doi{10.1016/j.nuclphysa.2005.03.085}{\bibinfo {journal} {Nucl. Phys.}}\ }%
  \textbf{\bibinfo {volume} {A757}},\ \bibinfo {pages} {102} (\bibinfo {year}
  {2005})%
  \bibAnnoteFile{NoStop}{Adams:2005dq}%
\bibitem{Adcox:2004mh}%
  \BibitemOpen
  \bibfield{author}{%
  \bibinfo {author} {\bibfnamefont{K.}~\bibnamefont{Adcox}} \emph{et~al.}
  (\bibinfo {collaboration} {PHENIX}),\ }%
  \bibfield{journal}{%
  \Doi{10.1016/j.nuclphysa.2005.03.086}{\bibinfo {journal} {Nucl. Phys.}}\ }%
  \textbf{\bibinfo {volume} {A757}},\ \bibinfo {pages} {184} (\bibinfo {year}
  {2005})%
  \bibAnnoteFile{NoStop}{Adcox:2004mh}%
\bibitem{Kolb:2003dz}%
  \BibitemOpen
  \bibfield{author}{%
  \bibinfo {author} {\bibfnamefont{P.~F.}\ \bibnamefont{Kolb}}\ and\ \bibinfo
  {author} {\bibfnamefont{U.~W.}\ \bibnamefont{Heinz}},\ }%
  in\ \emph{\bibinfo {booktitle} {Quark Gluon Plasma 3}},\ \bibinfo {editor}
  {edited by\ \bibinfo {editor} {\bibfnamefont{R.}~\bibnamefont{Hwa}}\ and\
  \bibinfo {editor} {\bibfnamefont{X.~N.}\ \bibnamefont{Wang}}}\ (\bibinfo
  {publisher} {World Scientific, Singapore},\ \bibinfo {year} {2004})\
  \Eprint{http://arxiv.org/abs/nucl-th/0305084}{arXiv:nucl-th/0305084}%
  \bibAnnoteFile{NoStop}{Kolb:2003dz}%
\bibitem{Huovinen:2006jp}%
  \BibitemOpen
  \bibfield{author}{%
  \bibinfo {author} {\bibfnamefont{P.}~\bibnamefont{Huovinen}}\ and\ \bibinfo
  {author} {\bibfnamefont{P.~V.}\ \bibnamefont{Ruuskanen}},\ }%
  \bibfield{journal}{%
  \Doi{10.1146/annurev.nucl.54.070103.181236}{\bibinfo {journal} {Ann. Rev.
  Nucl. Part. Sci.}}\ }%
  \textbf{\bibinfo {volume} {56}},\ \bibinfo {pages} {163} (\bibinfo {year}
  {2006})%
  \bibAnnoteFile{NoStop}{Huovinen:2006jp}%
\bibitem{Hirano:2008aj}%
  \BibitemOpen
  \bibfield{author}{%
  \bibinfo {author} {\bibfnamefont{T.}~\bibnamefont{Hirano}},\ }%
  \bibfield{journal}{%
  \Doi{10.1088/0954-3899/36/6/064031}{\bibinfo {journal} {J. Phys.}}\ }%
  \textbf{\bibinfo {volume} {G36}},\ \bibinfo {pages} {064031} (\bibinfo {year}
  {2009})%
  \bibAnnoteFile{NoStop}{Hirano:2008aj}%
\bibitem{Ollitrault:2010tn}%
  \BibitemOpen
  \bibfield{author}{%
  \bibinfo {author} {\bibfnamefont{J.-Y.}\ \bibnamefont{Ollitrault}}}%
   (\bibinfo {year} {2010}),\
  \Eprint{http://arxiv.org/abs/1008.3323}{arXiv:1008.3323 [nucl-th]}%
  \bibAnnoteFile{NoStop}{Ollitrault:2010tn}%
\bibitem{Broniowski:2008vp}%
  \BibitemOpen
  \bibfield{author}{%
  \bibinfo {author} {\bibfnamefont{W.}~\bibnamefont{Broniowski}}, \bibinfo
  {author} {\bibfnamefont{M.}~\bibnamefont{Chojnacki}}, \bibinfo {author}
  {\bibfnamefont{W.}~\bibnamefont{Florkowski}},\ and\ \bibinfo {author}
  {\bibfnamefont{A.}~\bibnamefont{Kisiel}},\ }%
  \bibfield{journal}{%
  \Doi{10.1103/PhysRevLett.101.022301}{\bibinfo {journal} {Phys. Rev. Lett.}}\
  }%
  \textbf{\bibinfo {volume} {101}},\ \bibinfo {pages} {022301} (\bibinfo {year}
  {2008})%
  \bibAnnoteFile{NoStop}{Broniowski:2008vp}%
\bibitem{Pratt:2008qv}%
  \BibitemOpen
  \bibfield{author}{%
  \bibinfo {author} {\bibfnamefont{S.}~\bibnamefont{Pratt}},\ }%
  \bibfield{journal}{%
  \Doi{10.1103/PhysRevLett.102.232301}{\bibinfo {journal} {Phys. Rev. Lett.}}\
  }%
  \textbf{\bibinfo {volume} {102}},\ \bibinfo {pages} {232301} (\bibinfo {year}
  {2009})%
  \bibAnnoteFile{NoStop}{Pratt:2008qv}%
\bibitem{Bozek:2009ty}%
  \BibitemOpen
  \bibfield{author}{%
  \bibinfo {author} {\bibfnamefont{P.}~\bibnamefont{Bo\.zek}}\ and\ \bibinfo
  {author} {\bibfnamefont{I.}~\bibnamefont{Wyskiel}},\ }%
  \bibfield{journal}{%
  \Doi{10.1103/PhysRevC.79.044916}{\bibinfo {journal} {Phys. Rev.}}\ }%
  \textbf{\bibinfo {volume} {C79}},\ \bibinfo {pages} {044916} (\bibinfo {year}
  {2009})%
  \bibAnnoteFile{NoStop}{Bozek:2009ty}%
\bibitem{IS}%
  \BibitemOpen
  \bibfield{author}{%
  \bibinfo {author} {\bibfnamefont{W.}~\bibnamefont{Israel}}\ and\ \bibinfo
  {author} {\bibfnamefont{J.}~\bibnamefont{Stewart}},\ }%
  \bibfield{journal}{%
  \bibinfo {journal} {Annals Phys.}\ }%
  \textbf{\bibinfo {volume} {118}},\ \bibinfo {pages} {341} (\bibinfo {year}
  {1979})%
  \bibAnnoteFile{NoStop}{IS}%
\bibitem{Teaney:2003kp}%
  \BibitemOpen
  \bibfield{author}{%
  \bibinfo {author} {\bibfnamefont{D.}~\bibnamefont{Teaney}},\ }%
  \bibfield{journal}{%
  \bibinfo {journal} {Phys. Rev.}\ }%
  \textbf{\bibinfo {volume} {C68}},\ \bibinfo {pages} {034913} (\bibinfo {year}
  {2003})%
  \bibAnnoteFile{NoStop}{Teaney:2003kp}%
\bibitem{Song:2007ux}%
  \BibitemOpen
  \bibfield{author}{%
  \bibinfo {author} {\bibfnamefont{H.}~\bibnamefont{Song}}\ and\ \bibinfo
  {author} {\bibfnamefont{U.~W.}\ \bibnamefont{Heinz}},\ }%
  \bibfield{journal}{%
  \Doi{10.1103/PhysRevC.77.064901}{\bibinfo {journal} {Phys. Rev.}}\ }%
  \textbf{\bibinfo {volume} {C77}},\ \bibinfo {pages} {064901} (\bibinfo {year}
  {2008})%
  \bibAnnoteFile{NoStop}{Song:2007ux}%
\bibitem{Dusling:2009df}%
  \BibitemOpen
  \bibfield{author}{%
  \bibinfo {author} {\bibfnamefont{K.}~\bibnamefont{Dusling}}, \bibinfo
  {author} {\bibfnamefont{G.~D.}\ \bibnamefont{Moore}},\ and\ \bibinfo {author}
  {\bibfnamefont{D.}~\bibnamefont{Teaney}},\ }%
  \bibfield{journal}{%
  \Doi{10.1103/PhysRevC.81.034907}{\bibinfo {journal} {Phys. Rev.}}\ }%
  \textbf{\bibinfo {volume} {C81}},\ \bibinfo {pages} {034907} (\bibinfo {year}
  {2010})%
  \bibAnnoteFile{NoStop}{Dusling:2009df}%
\bibitem{Chaudhuri:2006jd}%
  \BibitemOpen
  \bibfield{author}{%
  \bibinfo {author} {\bibfnamefont{A.~K.}\ \bibnamefont{Chaudhuri}},\ }%
  \bibfield{journal}{%
  \bibinfo {journal} {Phys. Rev.}\ }%
  \textbf{\bibinfo {volume} {C74}},\ \bibinfo {pages} {044904} (\bibinfo {year}
  {2006})%
  \bibAnnoteFile{NoStop}{Chaudhuri:2006jd}%
\bibitem{Dusling:2007gi}%
  \BibitemOpen
  \bibfield{author}{%
  \bibinfo {author} {\bibfnamefont{K.}~\bibnamefont{Dusling}}\ and\ \bibinfo
  {author} {\bibfnamefont{D.}~\bibnamefont{Teaney}},\ }%
  \bibfield{journal}{%
  \Doi{10.1103/PhysRevC.77.034905}{\bibinfo {journal} {Phys. Rev.}}\ }%
  \textbf{\bibinfo {volume} {C77}},\ \bibinfo {pages} {034905} (\bibinfo {year}
  {2008})%
  \bibAnnoteFile{NoStop}{Dusling:2007gi}%
\bibitem{Romatschke:2009im}%
  \BibitemOpen
  \bibfield{author}{%
  \bibinfo {author} {\bibfnamefont{P.}~\bibnamefont{Romatschke}},\ }%
  \bibfield{journal}{%
  \Doi{10.1142/S0218301310014613}{\bibinfo {journal} {Int. J. Mod. Phys.}}\ }%
  \textbf{\bibinfo {volume} {E19}},\ \bibinfo {pages} {1} (\bibinfo {year}
  {2010})%
  \bibAnnoteFile{NoStop}{Romatschke:2009im}%
\bibitem{Teaney:2009qa}%
  \BibitemOpen
  \bibfield{author}{%
  \bibinfo {author} {\bibfnamefont{D.~A.}\ \bibnamefont{Teaney}}}%
   (\bibinfo {year} {2009}),\
  \Eprint{http://arxiv.org/abs/0905.2433}{arXiv:0905.2433 [nucl-th]}%
  \bibAnnoteFile{NoStop}{Teaney:2009qa}%
\bibitem{Luzum:2008cw}%
  \BibitemOpen
  \bibfield{author}{%
  \bibinfo {author} {\bibfnamefont{M.}~\bibnamefont{Luzum}}\ and\ \bibinfo
  {author} {\bibfnamefont{P.}~\bibnamefont{Romatschke}},\ }%
  \bibfield{journal}{%
  \Doi{10.1103/PhysRevC.78.034915}{\bibinfo {journal} {Phys. Rev.}}\ }%
  \textbf{\bibinfo {volume} {C78}},\ \bibinfo {pages} {034915} (\bibinfo {year}
  {2008})%
  \bibAnnoteFile{NoStop}{Luzum:2008cw}%
\bibitem{Bozek:2009dw}%
  \BibitemOpen
  \bibfield{author}{%
  \bibinfo {author} {\bibfnamefont{P.}~\bibnamefont{Bo\.zek}},\ }%
  \bibfield{journal}{%
  \bibinfo {journal} {Phys. Rev.}\ }%
  \textbf{\bibinfo {volume} {C81}},\ \bibinfo {pages} {034909} (\bibinfo {year}
  {2010})%
  \bibAnnoteFile{NoStop}{Bozek:2009dw}%
\bibitem{Schenke:2010rr}%
  \BibitemOpen
  \bibfield{author}{%
  \bibinfo {author} {\bibfnamefont{B.}~\bibnamefont{Schenke}}, \bibinfo
  {author} {\bibfnamefont{S.}~\bibnamefont{Jeon}},\ and\ \bibinfo {author}
  {\bibfnamefont{C.}~\bibnamefont{Gale}}}%
   (\bibinfo {year} {2010}),\
  \Eprint{http://arxiv.org/abs/1009.3244}{arXiv:1009.3244 [hep-ph]}%
  \bibAnnoteFile{NoStop}{Schenke:2010rr}%
\bibitem{Mrowczynski:2005ki}%
  \BibitemOpen
  \bibfield{author}{%
  \bibinfo {author} {\bibfnamefont{S.}~\bibnamefont{Mrowczynski}},\ }%
  \bibfield{journal}{%
  \bibinfo {journal} {Acta Phys. Polon.}\ }%
  \textbf{\bibinfo {volume} {B37}},\ \bibinfo {pages} {427} (\bibinfo {year}
  {2006})%
  \bibAnnoteFile{NoStop}{Mrowczynski:2005ki}%
\bibitem{Rebhan:2008uj}%
  \BibitemOpen
  \bibfield{author}{%
  \bibinfo {author} {\bibfnamefont{A.}~\bibnamefont{Rebhan}}, \bibinfo {author}
  {\bibfnamefont{M.}~\bibnamefont{Strickland}},\ and\ \bibinfo {author}
  {\bibfnamefont{M.}~\bibnamefont{Attems}},\ }%
  \bibfield{journal}{%
  \Doi{10.1103/PhysRevD.78.045023}{\bibinfo {journal} {Phys. Rev.}}\ }%
  \textbf{\bibinfo {volume} {D78}},\ \bibinfo {pages} {045023} (\bibinfo {year}
  {2008})%
  \bibAnnoteFile{NoStop}{Rebhan:2008uj}%
\bibitem{Bjoraker:2000cf}%
  \BibitemOpen
  \bibfield{author}{%
  \bibinfo {author} {\bibfnamefont{J.}~\bibnamefont{Bjoraker}}\ and\ \bibinfo
  {author} {\bibfnamefont{R.}~\bibnamefont{Venugopalan}},\ }%
  \bibfield{journal}{%
  \Doi{10.1103/PhysRevC.63.024609}{\bibinfo {journal} {Phys. Rev.}}\ }%
  \textbf{\bibinfo {volume} {C63}},\ \bibinfo {pages} {024609} (\bibinfo {year}
  {2001})%
  \bibAnnoteFile{NoStop}{Bjoraker:2000cf}%
\bibitem{Xu:2004mz}%
  \BibitemOpen
  \bibfield{author}{%
  \bibinfo {author} {\bibfnamefont{Z.}~\bibnamefont{Xu}}\ and\ \bibinfo
  {author} {\bibfnamefont{C.}~\bibnamefont{Greiner}},\ }%
  \bibfield{journal}{%
  \Doi{10.1103/PhysRevC.71.064901}{\bibinfo {journal} {Phys. Rev.}}\ }%
  \textbf{\bibinfo {volume} {C71}},\ \bibinfo {pages} {064901} (\bibinfo {year}
  {2005})%
  \bibAnnoteFile{NoStop}{Xu:2004mz}%
\bibitem{Chesler:2009cy}%
  \BibitemOpen
  \bibfield{author}{%
  \bibinfo {author} {\bibfnamefont{P.~M.}\ \bibnamefont{Chesler}}\ and\
  \bibinfo {author} {\bibfnamefont{L.~G.}\ \bibnamefont{Yaffe}},\ }%
  \bibfield{journal}{%
  \Doi{10.1103/PhysRevD.82.026006}{\bibinfo {journal} {Phys. Rev.}}\ }%
  \textbf{\bibinfo {volume} {D82}},\ \bibinfo {pages} {026006} (\bibinfo {year}
  {2010})%
  \bibAnnoteFile{NoStop}{Chesler:2009cy}%
\bibitem{Beuf:2009cx}%
  \BibitemOpen
  \bibfield{author}{%
  \bibinfo {author} {\bibfnamefont{G.}~\bibnamefont{Beuf}}, \bibinfo {author}
  {\bibfnamefont{M.~P.}\ \bibnamefont{Heller}}, \bibinfo {author}
  {\bibfnamefont{R.~A.}\ \bibnamefont{Janik}},\ and\ \bibinfo {author}
  {\bibfnamefont{R.}~\bibnamefont{Peschanski}},\ }%
  \bibfield{journal}{%
  \Doi{10.1088/1126-6708/2009/10/043}{\bibinfo {journal} {JHEP}}\ }%
  \textbf{\bibinfo {volume} {10}},\ \bibinfo {pages} {043} (\bibinfo {year}
  {2009})%
  \bibAnnoteFile{NoStop}{Beuf:2009cx}%
\bibitem{Song:2008hj}%
  \BibitemOpen
  \bibfield{author}{%
  \bibinfo {author} {\bibfnamefont{H.}~\bibnamefont{Song}}\ and\ \bibinfo
  {author} {\bibfnamefont{U.~W.}\ \bibnamefont{Heinz}},\ }%
  \bibfield{journal}{%
  \Doi{10.1088/0954-3899/36/6/064033}{\bibinfo {journal} {J. Phys.}}\ }%
  \textbf{\bibinfo {volume} {G36}},\ \bibinfo {pages} {064033} (\bibinfo {year}
  {2009})%
  \bibAnnoteFile{NoStop}{Song:2008hj}%
\bibitem{Kovtun:2004de}%
  \BibitemOpen
  \bibfield{author}{%
  \bibinfo {author} {\bibfnamefont{P.~K.}\ \bibnamefont{Kovtun}}, \bibinfo
  {author} {\bibfnamefont{D.~T.}\ \bibnamefont{Son}},\ and\ \bibinfo {author}
  {\bibfnamefont{A.~O.}\ \bibnamefont{Starinets}},\ }%
  \bibfield{journal}{%
  \Doi{10.1103/PhysRevLett.94.111601}{\bibinfo {journal} {Phys. Rev. Lett.}}\
  }%
  \textbf{\bibinfo {volume} {94}},\ \bibinfo {pages} {111601} (\bibinfo {year}
  {2005})%
  \bibAnnoteFile{NoStop}{Kovtun:2004de}%
\bibitem{Danielewicz:1984ww}%
  \BibitemOpen
  \bibfield{author}{%
  \bibinfo {author} {\bibfnamefont{P.}~\bibnamefont{Danielewicz}}\ and\
  \bibinfo {author} {\bibfnamefont{M.}~\bibnamefont{Gyulassy}},\ }%
  \bibfield{journal}{%
  \Doi{10.1103/PhysRevD.31.53}{\bibinfo {journal} {Phys. Rev.}}\ }%
  \textbf{\bibinfo {volume} {D31}},\ \bibinfo {pages} {53} (\bibinfo {year}
  {1985})%
  \bibAnnoteFile{NoStop}{Danielewicz:1984ww}%
\bibitem{Liao:2009gb}%
  \BibitemOpen
  \bibfield{author}{%
  \bibinfo {author} {\bibfnamefont{J.}~\bibnamefont{Liao}}\ and\ \bibinfo
  {author} {\bibfnamefont{V.}~\bibnamefont{Koch}},\ }%
  \bibfield{journal}{%
  \Doi{10.1103/PhysRevC.81.014902}{\bibinfo {journal} {Phys. Rev.}}\ }%
  \textbf{\bibinfo {volume} {C81}},\ \bibinfo {pages} {014902} (\bibinfo {year}
  {2010})%
  \bibAnnoteFile{NoStop}{Liao:2009gb}%
\bibitem{Martinez:2010sc}%
  \BibitemOpen
  \bibfield{author}{%
  \bibinfo {author} {\bibfnamefont{M.}~\bibnamefont{Martinez}}\ and\ \bibinfo
  {author} {\bibfnamefont{M.}~\bibnamefont{Strickland}},\ }%
  \bibfield{journal}{%
  \Doi{10.1016/j.nuclphysa.2010.08.011}{\bibinfo {journal} {Nucl. Phys.}}\ }%
  \textbf{\bibinfo {volume} {A848}},\ \bibinfo {pages} {183} (\bibinfo {year}
  {2010})%
  \bibAnnoteFile{NoStop}{Martinez:2010sc}%
\bibitem{Martinez:2010sd}%
  \BibitemOpen
  \bibfield{author}{%
  \bibinfo {author} {\bibfnamefont{M.}~\bibnamefont{Martinez}}\ and\ \bibinfo
  {author} {\bibfnamefont{M.}~\bibnamefont{Strickland}}}%
   (\bibinfo {year} {2010}),\
  \Eprint{http://arxiv.org/abs/1011.3056}{arXiv:1011.3056 [nucl-th]}%
  \bibAnnoteFile{NoStop}{Martinez:2010sd}%
\bibitem{Dusling:2010rm}%
  \BibitemOpen
  \bibfield{author}{%
  \bibinfo {author} {\bibfnamefont{K.}~\bibnamefont{Dusling}}, \bibinfo
  {author} {\bibfnamefont{T.}~\bibnamefont{Epelbaum}}, \bibinfo {author}
  {\bibfnamefont{F.}~\bibnamefont{Gelis}},\ and\ \bibinfo {author}
  {\bibfnamefont{R.}~\bibnamefont{Venugopalan}}}%
   (\bibinfo {year} {2010}),\
  \Eprint{http://arxiv.org/abs/1009.4363}{arXiv:1009.4363 [hep-ph]}%
  \bibAnnoteFile{NoStop}{Dusling:2010rm}%
\bibitem{Vredevoogd:2008id}%
  \BibitemOpen
  \bibfield{author}{%
  \bibinfo {author} {\bibfnamefont{J.}~\bibnamefont{Vredevoogd}}\ and\ \bibinfo
  {author} {\bibfnamefont{S.}~\bibnamefont{Pratt}},\ }%
  \bibfield{journal}{%
  \Doi{10.1103/PhysRevC.79.044915}{\bibinfo {journal} {Phys. Rev.}}\ }%
  \textbf{\bibinfo {volume} {C79}},\ \bibinfo {pages} {044915} (\bibinfo {year}
  {2009})%
  \bibAnnoteFile{NoStop}{Vredevoogd:2008id}%
\bibitem{Mauricio:2007vz}%
  \BibitemOpen
  \bibfield{author}{%
  \bibinfo {author} {\bibfnamefont{M.}~\bibnamefont{Martinez}}\ and\ \bibinfo
  {author} {\bibfnamefont{M.}~\bibnamefont{Strickland}},\ }%
  \bibfield{journal}{%
  \bibinfo {journal} {Phys. Rev. Lett.}\ }%
  \textbf{\bibinfo {volume} {100}},\ \bibinfo {pages} {102301} (\bibinfo {year}
  {2008})%
  \bibAnnoteFile{NoStop}{Mauricio:2007vz}%
\bibitem{Schenke:2006yp}%
  \BibitemOpen
  \bibfield{author}{%
  \bibinfo {author} {\bibfnamefont{B.}~\bibnamefont{Schenke}}\ and\ \bibinfo
  {author} {\bibfnamefont{M.}~\bibnamefont{Strickland}},\ }%
  \bibfield{journal}{%
  \Doi{10.1103/PhysRevD.76.025023}{\bibinfo {journal} {Phys. Rev.}}\ }%
  \textbf{\bibinfo {volume} {D76}},\ \bibinfo {pages} {025023} (\bibinfo {year}
  {2007})%
  \bibAnnoteFile{NoStop}{Schenke:2006yp}%
\bibitem{Dusling:2008xj}%
  \BibitemOpen
  \bibfield{author}{%
  \bibinfo {author} {\bibfnamefont{K.}~\bibnamefont{Dusling}}\ and\ \bibinfo
  {author} {\bibfnamefont{S.}~\bibnamefont{Lin}},\ }%
  \bibfield{journal}{%
  \Doi{10.1016/j.nuclphysa.2008.06.007}{\bibinfo {journal} {Nucl. Phys.}}\ }%
  \textbf{\bibinfo {volume} {A809}},\ \bibinfo {pages} {246} (\bibinfo {year}
  {2008})%
  \bibAnnoteFile{NoStop}{Dusling:2008xj}%
\bibitem{Dusling:2009bc}%
  \BibitemOpen
  \bibfield{author}{%
  \bibinfo {author} {\bibfnamefont{K.}~\bibnamefont{Dusling}},\ }%
  \bibfield{journal}{%
  \Doi{10.1016/j.nuclphysa.2010.04.001}{\bibinfo {journal} {Nucl. Phys.}}\ }%
  \textbf{\bibinfo {volume} {A839}},\ \bibinfo {pages} {70} (\bibinfo {year}
  {2010})%
  \bibAnnoteFile{NoStop}{Dusling:2009bc}%
\bibitem{Bhattacharya:2008mv}%
  \BibitemOpen
  \bibfield{author}{%
  \bibinfo {author} {\bibfnamefont{L.}~\bibnamefont{Bhattacharya}}\ and\
  \bibinfo {author} {\bibfnamefont{P.}~\bibnamefont{Roy}},\ }%
  \bibfield{journal}{%
  \Doi{10.1103/PhysRevC.79.054910}{\bibinfo {journal} {Phys. Rev.}}\ }%
  \textbf{\bibinfo {volume} {C79}},\ \bibinfo {pages} {054910} (\bibinfo {year}
  {2009})%
  \bibAnnoteFile{NoStop}{Bhattacharya:2008mv}%
\bibitem{Bozek:2007di}%
  \BibitemOpen
  \bibfield{author}{%
  \bibinfo {author} {\bibfnamefont{P.}~\bibnamefont{Bo\.zek}},\ }%
  \bibfield{journal}{%
  \bibinfo {journal} {Acta Phys. Polon.}\ }%
  \textbf{\bibinfo {volume} {B39}},\ \bibinfo {pages} {1375} (\bibinfo {year}
  {2008})%
  \bibAnnoteFile{NoStop}{Bozek:2007di}%
\bibitem{Florkowski:2010cf}%
  \BibitemOpen
  \bibfield{author}{%
  \bibinfo {author} {\bibfnamefont{W.}~\bibnamefont{Florkowski}}\ and\ \bibinfo
  {author} {\bibfnamefont{R.}~\bibnamefont{Ryblewski}}}%
   (\bibinfo {year} {2010}),\
  \Eprint{http://arxiv.org/abs/1007.0130}{arXiv:1007.0130 [nucl-th]}%
  \bibAnnoteFile{NoStop}{Florkowski:2010cf}%
\bibitem{Ryblewski:2010bs}%
  \BibitemOpen
  \bibfield{author}{%
  \bibinfo {author} {\bibfnamefont{R.}~\bibnamefont{Ryblewski}}\ and\ \bibinfo
  {author} {\bibfnamefont{W.}~\bibnamefont{Florkowski}},\ }%
  \bibfield{journal}{%
  \Doi{10.1088/0954-3899/38/1/015104}{\bibinfo {journal} {J. Phys.}}\ }%
  \textbf{\bibinfo {volume} {G38}},\ \bibinfo {pages} {015104} (\bibinfo {year}
  {2011})%
  \bibAnnoteFile{NoStop}{Ryblewski:2010bs}%
\bibitem{Broniowski:2008qk}%
  \BibitemOpen
  \bibfield{author}{%
  \bibinfo {author} {\bibfnamefont{W.}~\bibnamefont{Broniowski}}, \bibinfo
  {author} {\bibfnamefont{W.}~\bibnamefont{Florkowski}}, \bibinfo {author}
  {\bibfnamefont{M.}~\bibnamefont{Chojnacki}},\ and\ \bibinfo {author}
  {\bibfnamefont{A.}~\bibnamefont{Kisiel}},\ }%
  \bibfield{journal}{%
  \Doi{10.1103/PhysRevC.80.034902}{\bibinfo {journal} {Phys. Rev.}}\ }%
  \textbf{\bibinfo {volume} {C80}},\ \bibinfo {pages} {034902} (\bibinfo {year}
  {2009})%
  \bibAnnoteFile{NoStop}{Broniowski:2008qk}%
\bibitem{Bozek:2007qt}%
  \BibitemOpen
  \bibfield{author}{%
  \bibinfo {author} {\bibfnamefont{P.}~\bibnamefont{Bo\.zek}},\ }%
  \bibfield{journal}{%
  \Doi{10.1103/PhysRevC.77.034911}{\bibinfo {journal} {Phys. Rev.}}\ }%
  \textbf{\bibinfo {volume} {C77}},\ \bibinfo {pages} {034911} (\bibinfo {year}
  {2008})%
  \bibAnnoteFile{NoStop}{Bozek:2007qt}%
\bibitem{Bialas:2004su}%
  \BibitemOpen
  \bibfield{author}{%
  \bibinfo {author} {\bibfnamefont{A.}~\bibnamefont{Bia\l{}as}}\ and\ \bibinfo
  {author} {\bibfnamefont{W.}~\bibnamefont{Czy\.z}},\ }%
  \bibfield{journal}{%
  \bibinfo {journal} {Acta Phys. Polon.}\ }%
  \textbf{\bibinfo {volume} {B36}},\ \bibinfo {pages} {905} (\bibinfo {year}
  {2005})%
  \bibAnnoteFile{NoStop}{Bialas:2004su}%
\bibitem{Adil:2005qn}%
  \BibitemOpen
  \bibfield{author}{%
  \bibinfo {author} {\bibfnamefont{A.}~\bibnamefont{Adil}}\ and\ \bibinfo
  {author} {\bibfnamefont{M.}~\bibnamefont{Gyulassy}},\ }%
  \bibfield{journal}{%
  \Doi{10.1103/PhysRevC.72.034907}{\bibinfo {journal} {Phys. Rev.}}\ }%
  \textbf{\bibinfo {volume} {C72}},\ \bibinfo {pages} {034907} (\bibinfo {year}
  {2005})%
  \bibAnnoteFile{NoStop}{Adil:2005qn}%
\bibitem{Bozek:2010bi}%
  \BibitemOpen
  \bibfield{author}{%
  \bibinfo {author} {\bibfnamefont{P.}~\bibnamefont{Bo\.zek}}\ and\ \bibinfo
  {author} {\bibfnamefont{I.}~\bibnamefont{Wyskiel}},\ }%
  \bibfield{journal}{%
  \Doi{10.1103/PhysRevC.81.054902}{\bibinfo {journal} {Phys. Rev.}}\ }%
  \textbf{\bibinfo {volume} {C81}},\ \bibinfo {pages} {054902} (\bibinfo {year}
  {2010})%
  \bibAnnoteFile{NoStop}{Bozek:2010bi}%
\bibitem{Abelev:2008jga}%
  \BibitemOpen
  \bibfield{author}{%
  \bibinfo {author} {\bibfnamefont{B.~I.}\ \bibnamefont{Abelev}} \emph{et~al.}
  (\bibinfo {collaboration} {STAR}),\ }%
  \bibfield{journal}{%
  \Doi{10.1103/PhysRevLett.101.252301}{\bibinfo {journal} {Phys. Rev. Lett.}}\
  }%
  \textbf{\bibinfo {volume} {101}},\ \bibinfo {pages} {252301} (\bibinfo {year}
  {2008})%
  \bibAnnoteFile{NoStop}{Abelev:2008jga}%
\bibitem{Adler:2003cb}%
  \BibitemOpen
  \bibfield{author}{%
  \bibinfo {author} {\bibfnamefont{S.~S.}\ \bibnamefont{Adler}} \emph{et~al.}
  (\bibinfo {collaboration} {PHENIX}),\ }%
  \bibfield{journal}{%
  \Doi{10.1103/PhysRevC.69.034909}{\bibinfo {journal} {Phys. Rev.}}\ }%
  \textbf{\bibinfo {volume} {C69}},\ \bibinfo {pages} {034909} (\bibinfo {year}
  {2004})%
  \bibAnnoteFile{NoStop}{Adler:2003cb}%
\bibitem{Kisiel:2005hn}%
  \BibitemOpen
  \bibfield{author}{%
  \bibinfo {author} {\bibfnamefont{A.}~\bibnamefont{Kisiel}}, \bibinfo {author}
  {\bibfnamefont{T.}~\bibnamefont{{Ta\l{}u\'c}}}, \bibinfo {author}
  {\bibfnamefont{W.}~\bibnamefont{Broniowski}},\ and\ \bibinfo {author}
  {\bibfnamefont{W.}~\bibnamefont{Florkowski}},\ }%
  \bibfield{journal}{%
  \bibinfo {journal} {Comput. Phys. Commun.}\ }%
  \textbf{\bibinfo {volume} {174}},\ \bibinfo {pages} {669} (\bibinfo {year}
  {2006})%
  \bibAnnoteFile{NoStop}{Kisiel:2005hn}%
\bibitem{Adams:2004yc}%
  \BibitemOpen
  \bibfield{author}{%
  \bibinfo {author} {\bibfnamefont{J.}~\bibnamefont{Adams}} \emph{et~al.}
  (\bibinfo {collaboration} {STAR}),\ }%
  \bibfield{journal}{%
  \bibinfo {journal} {Phys. Rev.}\ }%
  \textbf{\bibinfo {volume} {C71}},\ \bibinfo {pages} {044906} (\bibinfo {year}
  {2005})%
  \bibAnnoteFile{NoStop}{Adams:2004yc}%
\bibitem{Adare:2010ux}%
  \BibitemOpen
  \bibfield{author}{%
  \bibinfo {author} {\bibfnamefont{A.}~\bibnamefont{Adare}} \emph{et~al.}
  (\bibinfo {collaboration} {PHENIX}),\ }%
  \bibfield{journal}{%
  \Doi{10.1103/PhysRevLett.105.062301}{\bibinfo {journal} {Phys. Rev. Lett.}}\
  }%
  \textbf{\bibinfo {volume} {105}},\ \bibinfo {pages} {062301} (\bibinfo {year}
  {2010})%
  \bibAnnoteFile{NoStop}{Adare:2010ux}%
\bibitem{Hirano:2005xf}%
  \BibitemOpen
  \bibfield{author}{%
  \bibinfo {author} {\bibfnamefont{T.}~\bibnamefont{Hirano}}, \bibinfo {author}
  {\bibfnamefont{U.~W.}\ \bibnamefont{Heinz}}, \bibinfo {author}
  {\bibfnamefont{D.}~\bibnamefont{Kharzeev}}, \bibinfo {author}
  {\bibfnamefont{R.}~\bibnamefont{Lacey}},\ and\ \bibinfo {author}
  {\bibfnamefont{Y.}~\bibnamefont{Nara}},\ }%
  \bibfield{journal}{%
  \Doi{10.1016/j.physletb.2006.03.060}{\bibinfo {journal} {Phys. Lett.}}\ }%
  \textbf{\bibinfo {volume} {B636}},\ \bibinfo {pages} {299} (\bibinfo {year}
  {2006})%
  \bibAnnoteFile{NoStop}{Hirano:2005xf}%
\bibitem{Hirano:2007xd}%
  \BibitemOpen
  \bibfield{author}{%
  \bibinfo {author} {\bibfnamefont{T.}~\bibnamefont{Hirano}}, \bibinfo {author}
  {\bibfnamefont{U.~W.}\ \bibnamefont{Heinz}}, \bibinfo {author}
  {\bibfnamefont{D.}~\bibnamefont{Kharzeev}}, \bibinfo {author}
  {\bibfnamefont{R.}~\bibnamefont{Lacey}},\ and\ \bibinfo {author}
  {\bibfnamefont{Y.}~\bibnamefont{Nara}},\ }%
  \bibfield{journal}{%
  \bibinfo {journal} {J. Phys.}\ }%
  \textbf{\bibinfo {volume} {G34}},\ \bibinfo {pages} {S879} (\bibinfo {year}
  {2007})%
  \bibAnnoteFile{NoStop}{Hirano:2007xd}%
\bibitem{Heller:2007qt}%
  \BibitemOpen
  \bibfield{author}{%
  \bibinfo {author} {\bibfnamefont{M.~P.}\ \bibnamefont{Heller}}\ and\ \bibinfo
  {author} {\bibfnamefont{R.~A.}\ \bibnamefont{Janik}},\ }%
  \bibfield{journal}{%
  \Doi{10.1103/PhysRevD.76.025027}{\bibinfo {journal} {Phys. Rev.}}\ }%
  \textbf{\bibinfo {volume} {D76}},\ \bibinfo {pages} {025027} (\bibinfo {year}
  {2007})%
  \bibAnnoteFile{NoStop}{Heller:2007qt}%
\end{thebibliography}%

\end{document}